\documentstyle[prl,aps,twocolumn]{revtex}
\large  

\input epsf
\begin{document}
\draft \twocolumn[\hsize\textwidth
\columnwidth\hsize\csname
@twocolumnfalse\endcsname

\title{Critical dimensions of the diffusion equation} 
\author{T. J. Newman$^{(1)}$\cite{tea} and Will Loinaz$^{(2)}$\cite{wea}} 
\address{$^{(1)}$Departments of Physics and Biology, University of Virginia,
Charlottesville, VA 22904\\
$^{(2)}$Department of Physics, Amherst College, Amherst, MA 01002} 
\maketitle
\begin{abstract}
We study the evolution of a random initial field under pure
diffusion in various space dimensions. From numerical calculations we find
that the persistence properties of the system show sharp transitions
at critical dimensions $d_{1} \simeq 26$ and $d_{2} \simeq 46$. We also
give refined measurements of the persistence exponents for low dimensions.
\end{abstract}
\vspace{2mm} ]

\narrowtext

Diffusion is a ubiquitous process in Nature and is of fundamental
importance in physical, chemical, and biological systems. Pure
diffusion, as described by the simple heat equation, has been so
well-studied over so many years that it is difficult to imagine 
there is still much about it to be understood.  Recently, the
persistence probability, $q(t)$, of the diffusion equation in one, two, and
three dimensions was measured and found to be described by
non-trivial power laws\cite{depers}: $q(t) \sim t^{-\theta (d)}$, where
$d$ is the dimension of space. 
As yet there is no exact prediction for the exponents $\theta (d)$.

The study of persistence is a rapidly emerging field\cite{rev}, the
purpose of which is to probe the statistical nature of dynamical
processes with more resolution than that allowed by conventional
measures, such as autocorrelation functions. It has been applied
to a wide range of systems including phase ordering,
interface kinetics, and the diffusion equation. 
Direct experimental observation of persistence has been achieved
in the phase ordering of liquid crystals\cite{yurke} and just recently by
NMR studies of the diffusing polarization of a one dimensional 
gas\cite{diffexp}. Strictly speaking, the
persistence probability measures the probability that a
particular event has not occurred up to time $t$. More loosely,
by ``persistent properties'', we include other statistical
measures, such as the sign-time distribution
(STD)\cite{std1,std2}. The STD is a histogram measuring the fraction
of time $t$ a bimodal variable has been in one of its two
states. Examples are the fraction of time an Ising spin is ``up'', or
the fraction of time a stochastic variable is above its mean. The tails
of the STD implicitly encode the persistence probability. The shape of the
STD yields much useful information about the spatio-temporal structure of
the dynamical process. For example, it can be used to infer long-lived 
features in interfaces undergoing kinetic roughening\cite{std3}.

In this Letter we introduce a simple algorithm which allows us to
study the persistence properties of the diffusion equation in
arbitrary dimensions.  Our main result is that there exist {\it two}
critical dimensions in the behavior of the STD. We find that for
dimensions below $d_{1} \simeq 26$ the STD is purely concave (and thus
typical sign-time histories dominate $q(t)$), while for dimensions above 
$d_{2} \simeq 46$ the STD is purely convex (and thus $q(t)$ 
is controlled by rare sign-time histories). For dimensions in the range
$d_{1} < d < d_{2}$ the STD has more than one extremum and the
description of $q(t)$ in terms of typical and rare events is less
straightforward.

We consider a scalar field $\phi({\bf x},t)$ in a $d$-dimensional
space which is to be evolved under the diffusion equation:
\begin{equation}
\label{de}
\partial _{t}\phi = D\nabla ^{2} \phi \ ,
\end{equation}
where we have $\phi ({\bf x},t=0) = \psi({\bf x})$. The random field $\psi $ is
taken to have zero mean and to be Gaussian distributed with correlations 
$\langle \psi ({\bf x})\psi ({\bf x}') \rangle = \Delta \delta ^{d}({\bf x}-
{\bf x}') $. The results that follow are completely independent of 
the nature of the distribution (so long as it describes
short-range correlations) and the values of $D$ and $\Delta$ (so long
as they are positive definite). We define the persistence probability  
for this system as the probability that the diffusion field at a given point
in space (the origin, say) has never changed sign up to time $t$. The
sign-time $\tau (t)$ 
is similarly defined as the proportion of time $t$ the field at
the origin has been positive. The STD is the histogram constructed from 
the ensemble of sign-times collected from a large number of realizations.

The persistence probability $q(t)$ for the diffusion equation has been measured
previously from explicit simulations of a discrete diffusion process on
a finite $d$-dimensional grid\cite{depers} for dimensions $d$=1, 2, and 3.
For a given system one collects the persistence time for each site
of the lattice and forms $q(t)$ by binning these values. There is an 
implicit assumption concerning self-averaging since these persistence
times are not independent. This is equivalent to stating that 
$q(t)$ also measures the fraction of persistent sites in the system, which 
is certainly plausible in the large time limit.

We have measured $q(t)$ and the STD using an alternative algorithm which 
has many advantages over direct simulation of the discrete diffusion process.
The algorithm is constructed as follows. First we note that the diffusion 
equation may be explicitly integrated. Restricting our attention to the 
diffusion field at the origin we have
\begin{equation}
\label{desol1}
\phi ({\bf 0},t) = \int d^{d}y \ g({\bf y},t) \psi ({\bf y}) \ ,
\end{equation}
where $g({\bf x},t) = (4\pi Dt)^{-d/2}\exp (-x^{2}/4Dt)$ is the heat
kernel. The key observation is that the solid angle integration in
Eq.(\ref{desol1}) may be absorbed into a redefinition of the random field,
as originally noted by Hilhorst\cite{hil}. 
Defining a Gaussian random field $\Psi (r)$
with correlations $\langle \Psi (r) \Psi (r') \rangle = \delta (r-r')$, we 
have
\begin{equation}
\label{desol2}
\phi ({\bf 0},t) \propto \int \limits _{0}^{\infty} dr \ r^{(d-1)/2} 
e^{-r^{2}/t} \Psi (r) \ ,
\end{equation}
where we have scaled out $D$ and $\Delta $ 
and ignored the (positive definite) prefactor, 
as only the sign of $\phi ({\bf 0},t)$ is relevant to the persistent 
properties. To measure $q(t)$ for a given dimension $d$ we construct
a random field $\Psi(r)$ and perform the radial integral above for a series of
incremented times, until the integral changes sign. We record the time at
which this sign change occurs and repeat the procedure for another realization
of $\Psi(r)$. This collection of persistence times is then used to construct
$q(t)$. If we wish to measure the STD we evaluate the integral for
a series of incremented times within a fixed time range and record the fraction
of time for which the integral is positive. The computer time required for 
these procedures is essentially independent of the value of $d$.

This algorithm is superior to direct numerical simulation in many respects:
i) any dimension may be studied (even non-integer), ii) 
there are no finite size effects, iii) there is no implicit assumption
concerning self-averaging, iv) in measuring $q(t)$ 
a given realization is discarded as soon
as the field changes sign (in a direct simulation, all field points
are studied even if the vast majority are no longer persistent), v) the
algorithm may be optimized allowing very long time intervals to be studied
with ease.

We shall now discuss a few technical details 
concerning optimization.  One may think that in order to measure
$q(t)$ one must evaluate the integral in (\ref{desol2}) on a fine
grid of incremented times, so as not to miss any event in which the
field changes sign twice over a short time interval. In fact this is
not the case. For a given realization of $\Psi$ we have explicitly
evaluated the function $\phi({\bf 0},t)$ on a fine mesh of time slices
and find it is smooth, becoming more so as time proceeds. This
is reasonable when one considers that under diffusion the field at a
given point will have increasingly smoother dynamics with time.
Due to the progressive smoothness of $\phi({\bf 0},t)$ 
we need only sample it on a logarithmic time scale. 

The base of the logarithm to be used is determined from calibration. We find 
the exact zeroes of $10^{4}$ realizations over three decades of time using
a fine linear mesh of time slices.  We then run the optimized code
which searches for these zeroes first using a logarithmic scale, and
then using recursive bisection of a logarithmic interval if it has reason to
believe a zero lies within. The logarithm base is chosen so that 
{\it all} of the zeroes from the $10^{4}$ samples are found
correctly. This limits the systematic error of our results to less
than 0.01\%. The core of the algorithm concerns the detection of
zeroes within a logarithmic interval, based on the value of the
function $\phi ({\bf 0},t)$ at the boundaries of the interval. We use
two techniques. The faster, but less precise, algorithm uses recursive 
bisection of an interval only if the values of $\phi ({\bf 0},t)$ at the
boundaries have different signs (in which case there
has to be an odd number of zeroes in the interval). 
The more sophisticated algorithm
also uses information about the derivatives of $\phi ({\bf 0},t)$ at
the boundaries of the interval. This algorithm will also use recursive 
bisection to check for a zero if the values of $\phi ({\bf 0},t)$ at the
boundaries have the same sign, but the derivatives
have opposite signs (indicating an extremum in the interval and the possibility
of an even number of zeroes).
As an example of calibration, for $d=1$ the first algorithm finds all zeroes
correctly over three time decades if the logarithm base used is
1.1. The second algorithm finds all zeroes correctly with the
logarithm base 2.0. (As $d$ increases the second algorithm requires
a smaller logarithmic base to find all the zeroes -- as an example, for $d=20$
the required base is 1.2.) For measurements of $q(t)$ we always use the
second algorithm, but for measuring the STD we generally use the first
algorithm (as the STD requires much longer computing time).

Using this method we have been able to measure $q(t)$ and the STD for a 
wide range of dimensions. For dimensions $d$=1, 2, and 3 we are able to make 
estimates for the persistence exponent $\theta$ as shown in Table 1,
with much higher precision than formerly obtained\cite{depers}. In Table 1
we also give estimates for $\theta $ in higher dimensions, which 
until now have not been measured. By comparing these results with those
obtained from the independent interval approximation (IIA) \cite{depers}
we see that the IIA consistently underestimates the exponent values. 

\begin{center}
\begin{tabular}{|c|c||c|c|}\hline 
{ \ \ \ $d$ \ \ \ } & { \ \ \ \ $\theta (d)$ \ \ \ \ } & { \ \ \ $d$ \ \ \ } & 
{ \ \ \ \ $\theta (d)$ \ \ \ \ } \\ \hline 
  1.0 & \hspace{0.5mm} 0.12050(5)  & 10.0 & 0.4587(2) \\ \hline
  2.0 & 0.1875(1)  & 20.0 & 0.6556(2) \\ \hline
  3.0 & 0.2382(1)  & 30.0 & 0.8053(3) \\ \hline
  4.0 & 0.2806(2)  & 40.0 & 0.9312(4) \\ \hline
  5.0 & 0.3173(2)  & 50.0 & 1.0415(5) \\ \hline \hline
 26.0 & 0.7491(3)  & 46.0 & 1.0010(5) \\ \hline
\end{tabular}
\end{center}

\noindent
Table 1: Numerically calculated values of the persistence exponent for a
range of dimensions. 

\vspace{0.2cm}

In Figures 1 and 2 we present log-log plots
of $q(t)$ for a sample of dimensions to show the quality of our data.
We generally calculate $\phi ({\bf 0},t)$ over six decades of time, and
average over $10^{8}$ realizations. 
Our exponent estimates were made in the same fashion as in Ref.\cite{depers},
namely, by plotting $t^{\theta }q(t)$ versus $\log t$ and adjusting 
the exponent until the data is as level as possible. The error bars are
estimated from the resolution of this procedure.

For the diffusion equation it is known that for large $d$ the 
persistence exponent diverges as $\sqrt{d}$\cite{depers}. By plotting our
measured exponents on a log-log scale we have fitted this behavior and find
$\theta \sim c\sqrt{d}$ with $c=0.1475(10)$.
At the bottom of Table 1 we have given exponent estimates for the two
critical dimensions in this system. To clarify the role of these
dimensions we shall now discuss our data for the STD's.

\begin{figure}[htbp]
\epsfxsize=3.0in 
\hspace*{-0.1cm}
\epsfbox{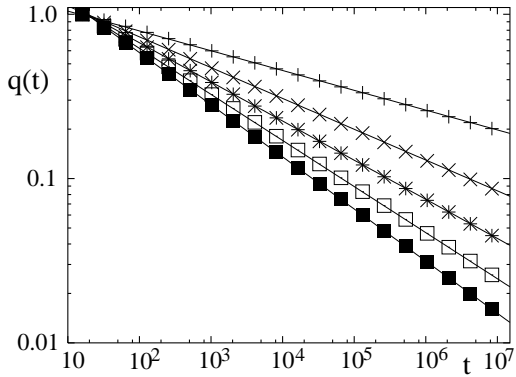}
\vspace*{-0.2cm}
\caption{Persistence probability $q(t)$ versus $t$, for dimensions
(from top to bottom) 1, 2, 3, 4, and 5.}
\end{figure}

\begin{figure}[htbp]
\epsfxsize=3.0in 
\vspace*{-0.5cm}
\hspace*{-0.1cm}
\epsfbox{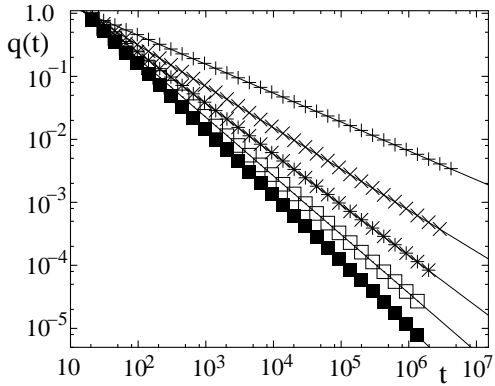}
\vspace*{-0.2cm}
\caption{Persistence probability $q(t)$ versus $t$, for dimensions
(from top to bottom) 10, 20, 30, 40, and 50.}
\end{figure}

To measure the STD, $P(\tau,t)$,
we evolve the system for three decades of time and by
locating zeroes of $\phi ({\bf 0},t)$ determine the sign-time $\tau(t)$
for a particular
realization. This process is repeated for $10^{8}$ samples and the STD is
constructed by binning the sign-times. In Ref.\cite{std2} it is shown
that the STD has the exact scaling form $P(\tau, t)=(1/t)f(u)$ where
$u=\tau/t$. In what follows we shall describe the STD in terms of the scaling
function $f(u)$, which is symmetric about $u=1/2$. 
The tails of the STD encode information about the persistence 
probability\cite{std2}. For example, the left tail is expected to vary as 
$u ^{\theta -1}$. Thus, for $\theta < 1$ the STD will have integrably 
divergent tails, whereas for $\theta > 1$ the STD will have vanishing tails.
Since $\theta $ diverges with $d$, there exists a critical
dimension $d_{2}$ at which $\theta (d_{2})=1$, and which separates (crudely
speaking) convex and concave STD's. 
From our numerical work we find that this scenario is too simple, and that
for a range of dimensions the STD has more than one extremum (as shown in 
Figure 3).

\begin{figure}[htbp]
\epsfxsize=3.0in 
\vspace*{-0.2cm}
\hspace*{-0.2cm}
\epsfbox{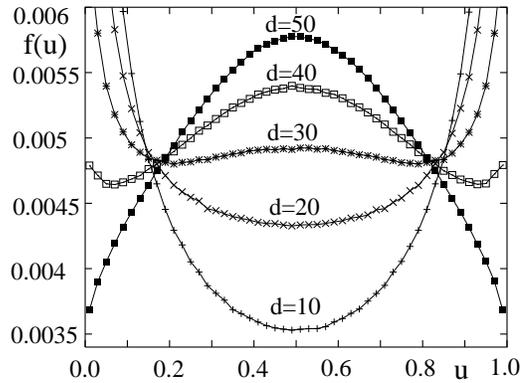}
\vspace*{-0.2cm}
\caption{ \ STD's for dimensions 10, 20, 30, 40, and 50.}
\end{figure}

We start by numerically determining $d_{2}$. By simply measuring the
persistence probability $q(t)$ for a range of dimensions we can identify that 
dimension at which $\theta =1$. As can be seen from Table 1, we find that 
$\theta (46.0) = 1.0010(5)$. A closer analysis leads us to the result
$d_{2}=45.9(1)$. This may be verified (to a lesser precision) by studying 
the STD's. In Figure 4 we show the tails of the STD's for $d=45.0$ and 
$d=46.0$. Note in the former case the upturn signalling a (very) 
weak integrable singularity in the distribution. 

\begin{figure}[htbp]
\epsfxsize=3.0in 
\vspace*{-0.3cm}
\hspace*{-0.1cm}
\epsfbox{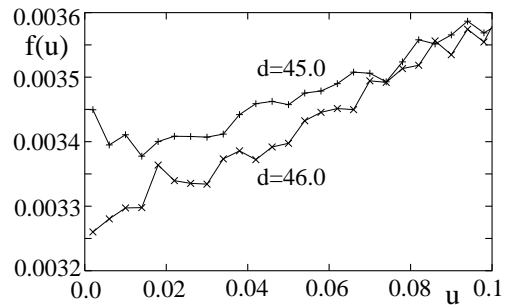}
\vspace*{-0.3cm}
\caption{Tails of the STD for $d=45.0$ and $d=46.0$.}
\end{figure}

From the STD's in Figure 3 we see there is another critical dimension $d_{1}$
separating STD's which are purely concave from those with three extrema.
There is no obvious way to identify this by measuring the persistence exponent.
Instead we must use direct measurements of the STD's which naturally leads
to a less precise estimate. In Figure 5 we show the measured STD's for
a small range of dimensions around $d=26.0$. From this data we arrive at
the result $d_{1}=26.0(5)$. In an earlier work\cite{std2} $d_{2}$
was estimated by assuming that the STD in $d_{2}$ dimensions is perfectly
flat. This led to the prediction $d_{2} = 35.967\dots $, which, within the
error bars, is the arithmetic mean of the measured values of $d_{1}$ and 
$d_{2}$.

Our main result with regard to the persistence properties of the diffusion 
equation
concerns the non-trivial behavior of the STD as a function of 
$d$. For $d<d_{1}=26.0(5)$ the STD is a purely concave function, and thus the
persistence probability (which is determined by the tails of the STD) is
dominated by typical sign-time histories. For $d>d_{2}=45.9(1)$ the
STD is a purely convex function, and in this case the persistence probability
is determined from rare events in the set of sign-time histories. There
is a range of dimensions $d_{1}<d<d_{2}$ for which the STD has three extrema
(as shown in Figure 3). In this case the persistence probability is still
described by the most typical sign-time histories, but we see that sign-time
histories around $u=1/2$ (meaning the field $\phi({\bf 0},t)$ spends half of 
its time above the mean) are relatively typical also, with the least typical
events occurring for some $d$-dependent value of $u$. These results 
are universal. The STD is independent of details
of the initial distribution of the diffusion field, and the value of the
diffusion constant. It is also asymptotically (i.e. for large times)
independent of higher-order processes, such as the corrections due to
an underlying lattice.

\begin{figure}[htbp]
\epsfxsize=3.0in 
\vspace*{-0.2cm}
\hspace*{-0.1cm}
\epsfbox{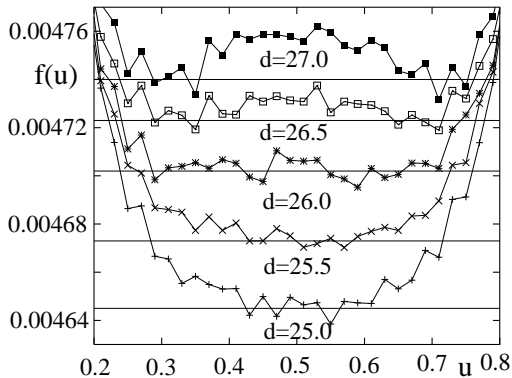}
\caption{STD's for a small range of dimensions about $d_{1}\simeq 26.0$.
The horizontal lines are a guide to the eye.}
\end{figure}

Clearly if one is to find applications of our high-dimensional
results, then the space
in which the diffusion process occurs cannot be taken as the familiar
physical space (at least within the realm of classical physics). A
very common manifestation of the diffusion equation is within the field
of stochastic processes. Indeed, the one dimensional 
diffusion equation is the simplest
possible form of the Fokker-Planck equation, describing the probability
distribution of a single random walker\cite{gard}. For $N$ independent
random walkers in one dimension, 
the multi-variate probability distribution $\rho (x_{1},\dots,x_{N})$ satisfies
the $N$-dimensional diffusion equation. Given random initial positions of the 
walkers, we should focus on the sign-time histories of $\rho$ with respect 
to its mean value over the ensemble of initial conditions. 
The distribution of these sign times will have sharp changes in behavior 
as the number of walkers $N$ is varied through $N_{1}=26$ and $N_{2}=46$.
If each walker exists in a ${\tilde d}$ dimensional space, then the
effective dimensionality of the diffusion equation is $N{\tilde d}$ and
the critical walker numbers $N_{1}$ and $N_{2}$ will be changed accordingly. 
Analogous statements may
be made for a system of $N$ directed polymers (e.g. magnetic flux lines) in
a ${\tilde d}$ dimensional space,
since the multi-variate partition function for $N$ lines also satisfies
the $N{\tilde d}$ dimensional diffusion equation\cite{poly}.

Our results also indicate that persistent fluctuations in the phase space
of few-body systems
may have sharp transitions on varying the number of degrees of freedom.
By ``few-body'' systems, we have in mind organic molecules composed of
tens of atoms, the dynamical properties of which are currently receiving a 
great deal of attention\cite{mole}.

In this Letter we have presented results for the persistence probability
and sign-time distribution for the diffusion equation over a wide range
of dimensions. These results have been obtained via a new algorithm which
allows extremely precise measurements of the persistence properties. This
algorithm is based on the integrability of the diffusion equation, a feature
shared by many other interesting problems such as the Edwards-Wilkinson model
of interface growth\cite{ew}, and the Burgers equation\cite{burg} of fluid
turbulence. The persistence
properties of these and similar models may be analyzed to high precision
using the methods outlined here.

\end{document}